\documentclass[sigconf]{acmart}
\usepackage{subfigure}
\usepackage{multirow}
\usepackage{booktabs}
\usepackage{color}
\usepackage{booktabs}

\AtBeginDocument{%
  \providecommand\BibTeX{{%
    \normalfont B\kern-0.5em{\scshape i\kern-0.25em b}\kern-0.8em\TeX}}}

\newtheoremstyle{mystyle}% % Name
    {1.0mm}%	% Space above
    {1.0mm}% 	% Space below
    {\it}%		% Body font
    {0mm}%	% Indent amount
    {\scshape}% % Theorem head font
    {.}%	% Punctuation after theorem head
    { }%	% Space after theorem head, ' ', or \newline
    {}%		% Theorem head spec (can be left empty, meaning `normal')
\theoremstyle{mystyle}
\newtheorem{defi}{Definition}
\newtheorem{exa}{Example}

\newcommand{\vsp}{\vspace{1.0mm}}

\setcopyright{acmcopyright}
\copyrightyear{2018}
\acmYear{2018}
\acmDOI{10.1145/1122445.1122456}

\setcopyright{acmcopyright}
\copyrightyear{}
\acmYear{}
\acmDOI{}
\acmConference[]{}{}{}
\acmBooktitle{}
\acmPrice{}
\acmISBN{}

\setcopyright{none}
\renewcommand\footnotetextcopyrightpermission[1]{} % removes footnote with conference information in first column
\begin{document}
\fancyhead{}

%%
%% The "title" command has an optional parameter,
%% allowing the author to define a "short title" to be used in page headers.
\title{Learned k-NN Distance Estimation}

%% The "author" command and its associated commands are used to define
%% the authors and their affiliations.
%% Of note is the shared affiliation of the first two authors, and the
%% "authornote" and "authornotemark" commands
%% used to denote shared contribution to the research.
\author{Daichi Amagata}
\authornote{Both authors contributed equally to this research.}
\affiliation{%
  \institution{Osaka University}
  \country{Japan}
}
\email{amagata.daichi@ist.osaka-u.ac.jp}

\author{Yusuke Arai}
\authornotemark[1]
\affiliation{%
  \institution{Osaka University}
  \country{Japan}
}
\email{arai.yusuke@ist.osaka-u.ac.jp}

\author{Sumio Fujita}
\affiliation{%
 \institution{Yahoo Japan Corporation}
 \country{Japan}}
\email{sufujita@yahoo-corp.jp }

\author{Takahiro Hara}
\affiliation{%
 \institution{Osaka University}
 \country{Japan}}
\email{hara@ist.osaka-u.ac.jp}

%%
%% By default, the full list of authors will be used in the page
%% headers. Often, this list is too long, and will overlap
%% other information printed in the page headers. This command allows
%% the author to define a more concise list
%% of authors' names for this purpose.
\renewcommand{\shortauthors}{Amagata et al.}

%%
%% The abstract is a short summary of the work to be presented in the
%% article.
\begin{abstract}
Big data mining is well known to be an important task for data science, because it can provide useful observations and new knowledge hidden in given large datasets.
Proximity-based data analysis is particularly utilized in many real-life applications.
In such analysis, the distances to $k$ nearest neighbors are usually employed, thus its main bottleneck is derived from data retrieval.
Much efforts have been made to improve the efficiency of these analyses.
However, they still incur large costs, because they essentially need many data accesses.
To avoid this issue, we propose a machine-learning technique that quickly and accurately estimates the $k$-NN distances (i.e., distances to the $k$ nearest neighbors) of a given query.
We train a fully connected neural network model and utilize pivots to achieve accurate estimation.
Our model is designed to have useful advantages: it infers distances to the $k$-NNs at a time, its inference time is $O(1)$ (no data accesses are incurred), but it keeps high accuracy.
Our experimental results and case studies on real datasets demonstrate the efficiency and effectiveness of our solution.
\end{abstract}

\maketitle

\section{Introduction}  \label{section_introduction}
Due to the proliferation of IoT devices, it is becoming easier to maintain big data.
Big data mining is well known to be an important task for data science, because it can provide useful observations and new knowledge hidden in given large datasets.
Proximity-based data analysis is particularly utilized in many real-life applications \cite{boukerche2020outlier, marcus2020benchmarking}.

\vsp
\noindent
\textbf{Motivation.}
Many proximity-based analysis tools employ distances to $k$ nearest neighbors ($k$-NNs), because they are an intuitive and effective way to measure the density (or sparsity) of a given point.
For example, geometric inference \cite{biau2011weighted}, outlier/anomaly detection \cite{amagata2021fast,amagata2022fast}, clustering \cite{campello2015hierarchical}, and cell analysis \cite{ziegler2020sars} employ this approach.
Below, by using some representative analysis tools that are used by industrial applications, we introduce how to use the distances to $k$-NNs more concretely.

\begin{exa}[\textsc{$k$-NN density estimation \& visualization}]  \label{example_density-estimation}
Density estimation \cite{loftsgaarden1965nonparametric} is one of the data visualization tools used most frequently, e.g., for detecting hot spots w.r.t. traffic and crime.
Given a query point $q$, its density based on $k$-NNs is obtained via the following estimator \cite{biau2011weighted}:
\begin{equation}
    \hat{e}(q) = \frac{1}{nV_{d}}\left(\frac{\sum_{j = 1}^{k}j^{d/2}}{\sum_{j=1}^{k}dist(q,x^{j}_{q})^d}\right)^{d/2},  \label{equation_knnd}
\end{equation}
where $n$ is the dataset size, $V_{d}$ is the volume of the unit hypersphere in a $d$-dimensional Euclidean space, $x^{j}_{q}$ is the $j$-th NN of $q$, and $dist(\cdot,\cdot)$ is the Euclidean distance between two points.
Given a space for visualization and its resolution, each pixel can be a query $q$ and obtains its $k$-NN density via the above equation, which is then displayed in the corresponding (sub-)space.
(The visualized result of a crime position dataset may be used by security service companies/organizations to highlight danger areas.)
\end{exa}

\begin{exa}[\textsc{Distance-based outlier detection}]  \label{example_dod}
The importance of outlier detection and its wide range of applications are well known, e.g., see survey papers \cite{boukerche2020outlier, chandola2009anomaly, wang2019progress}.
Let us consider machine learning (ML) models, and they are now used to automate operations in many companies.
However, ML models suffer from data errors \cite{schelter2021jenga}, and such errors (outliers) should be detected and removed to train high performance models.
Distance-based outlier detection is often utilized to achieve this, because it is intuitive and effective \cite{amagata2021fast}.
Literature \cite{knox1998algorithms} proposed a definition that a point $x$ is an outlier iff the distance between $x$ and its $k$-NN is larger than a threshold.
Similarly, literature \cite{ramaswamy2000efficient} proposed a definition that a point $x$ is an outlier iff the distance between $x$ and its $k$-NN is in the top-$N$ among a given dataset.
\end{exa}

\begin{exa}[\textsc{Reverse engineering for clustering}]    \label{example_dpc}
Let us consider density-peaks clustering (DPC) \cite{rodriguez2014clustering}, one of the density-based clustering algorithms.
DPC has been employed in many domains \cite{amagata2021dpc}.
For example, \cite{zhang2017incremental} used DPC to analyze industrial IoT datasets. 
As a criterion for clustering, DPC computes the local density $\rho_{i}$ for each point $x_{i}$, and it is the number of points $x_{j}$ such that $dist(x_{i},x_{j}) \leq d_{cut}$, where $d_{cut}$ is a threshold.
If $\rho_{i} < \rho_{min}$, DPC regards $x_{i}$ as noise, similarly to the distance-based outlier.

Assume that a result of DPC is given but the value of $d_{cut}$ is unknown.
Assume furthermore that we know the number of noises (from clustering labels) and $\rho_{min}$.
To reproduce the clustering result, we need to know $d_{cut}$, but, because $d_{cut}$ is a real value, testing all possible values is not practical.
Then, it is important to notice that, by measuring the distance to $\rho_{min}$-NN, we can see the distance that distinguishes noises.
(Details will be given in Section \ref{section_case-study-dpc}.)
From this observation, we can estimate $d_{cut}$.
\end{exa}

\noindent
The above examples clarify the importance of obtaining the \textit{distances to $k$-NNs}.

A straightforward approach to computing the $k$-NN distances is to retrieve the $k$-NN points.
Because the above applications face large datasets, there are many approaches that efficiently retrieve the $k$-NN points, such as tree indices \cite{bentley1975multidimensional, beygelzimer2006cover, qi2018theoretically}.
However, when the objective is to know the \textit{distances to $k$-NNs}, data retrieval approaches still incur large computational cost, because they need to \textit{access} many points.

\vsp
\noindent
\textbf{Contribution.}
In this paper, we consider a machine-learning (ML) technique to alleviate this cost.
Recently, ML techniques have been introduced in many database management systems \cite{dutt2019selectivity, marcus2020benchmarking, sun2019end, yang2019deep} because of their adaptability to the distributions of real datasets and accurate inference ability.
It is important to note that, to employ an ML technique for the problem of $k$-NN distance estimation, \textbf{the ML model has to be simple (to enable fast inference) yet accurate}.
We tackle this challenge and propose a regression-based solution.

Assume that we have a set $X$ of $d$-dimensional points, there are a lot of pivot points in the data space, and the distances to the $k$-NNs of the pivots are pre-computed.
Given a query point $q$, there would exist a pivot that is very near $q$.
Therefore, by using the $k$-NN distances of the pivot, we can infer the distances to the $k$-NNs of $q$ and its error would be very small.
It is however not feasible, in practice, to prepare pivots so that there exist sufficiently close pivots for arbitrary queries, as this approach is prohibitive w.r.t. space cost.
Hence, we train a regression model based on a fully connected neural network and exploit a reasonable number of pivots to learn an accurate model.
Our model, namely \textit{PivNet}, is designed to deal with multiple values of $k$ in $O(1)$ time, which is required in some applications such as Example \ref{example_density-estimation}.
In addition to the above application examples, our $k$-NN distance estimation enables to obtain a threshold for $k$-NN search.
This speeds-up the $k$-NN query processing because we do not need to set $\infty$ as the initial threshold any more (although the result may be approximate).
To summarize, our main contributions are as follows:
\begin{itemize}
    \setlength{\leftskip}{-4.0mm}
    \item   We propose PivNet, a learned model that infers the $k$-NN distances of a given query in $O(1)$ time.
            Its source code is open-source at \textcolor{blue}{\url{https://github.com/arailly/pivnet}}.
    \item   We conduct experiments to demonstrate that PivNet provides a fast inference time while keeping small errors.
    \item   We conduct case studies w.r.t. $k$-NN density visualization, distance-based outlier detection, reverse engineering for density-based clustering, and approximate $k$-NN search.
            The results show the practical efficiency and effectiveness of PivNet.
\end{itemize}
This is a full version of \cite{learned2022amagata}.

\vsp
\noindent
\textbf{Organization.}
The remainder of this paper is organized as follows.
Section \ref{section_preliminary} formally defines our problem.
Section \ref{section_solution} presents our solution.
Sections \ref{section_experiment} and \ref{section_case-study} respectively report our experimental and case studies.
Finally, Section \ref{section_conclusion} concludes this paper.

\section{Preliminary}   \label{section_preliminary}
Let $X$ be a set of $d$-dimensional points in the Euclidean space.
We use $dist(x,x')$ to denote the Euclidean distance between two points $x$ and $x'$.
Below, we define the $k$ nearest neighbors ($k$-NNs).

\begin{defi}[\textsc{$k$ nearest neighbors}]    \label{definition_knn}
Given a set $X$ of points, a query point $q$, and $k$, $A$ is a set of the $k$ nearest neighbors of $q$ that satisfies $|A| = k$ and $\forall x \in A, \forall x' \in X\backslash A, dist(x,q) \leq dist(x',q)$.
\end{defi}

\noindent
Then, our problem is defined as follows:

\begin{defi}[$k$-NN distance estimation problem]    \label{definition_problem}
Given a set $X$ of points, a query $q$, and $k$, let $A$ be a set of the $k$ nearest neighbors of $q$ defined in Definition \ref{definition_knn}.
This problem estimates the distance between $q$ and $x$ for each $x \in A$.
\end{defi}

We assume that $X$ is static (or is not frequently updated), which is a common assumption, e.g., for applications in Examples \ref{example_density-estimation}--\ref{example_dpc}.
Furthermore, we assume that there is a maximum $k$, denoted by $k_{max}$ (e.g., $k_{max}$ is 50 or 100), which is specified by applications (so that it satisfies their requirements).
Our solution is then designed to estimate the distances to $k_{max}$-NNs.

Consider a query $q$, and let $\hat{dist}(q,x_{q}^{k})$ be an estimated distance to its $k$-NN.
For distance estimation, minimizing absolute error $|dist(q,x_{q}^{k}) - \hat{dist}(q,x_{q}^{k})|$ is clearly important, as can be seen from Examples \ref{example_dod} and \ref{example_dpc}.
Therefore, as an error metric, we mainly use MAE (Mean Absolute Error), and we compute the MAE of $q$, $MAE_{q}$, as follows:
\begin{equation}
    MAE_{q} = \frac{\sum_{k = 1}^{k_{max}}|dist(q,x_{q}^{k}) - \hat{dist}(q,x_{q}^{k})|}{k_{max}}.  \label{equation_mae}
\end{equation}
(In our empirical evaluation, Mean Absolute Percentage Error, MAPE, is also measured.)

Because the application examples in Section \ref{section_introduction} typically assume low-dimensional datasets \cite{amagata2021dpc, cao2017multi, chan2020quad, chan2021fast, wang2020theoretically}, this paper considers that $d$ is low.
The case of high-dimensional data is a future work, because solutions for this case can be totally different\footnote{This is trivial.
For example, space partitioning methods, such as tree indices, are generally used for similarity search on low-dimensional data, whereas they are not used for high-dimensional data \cite{li2019approximate}.}.

\begin{figure*}[!t]
	\begin{center}
	    \includegraphics[width=0.90\linewidth]{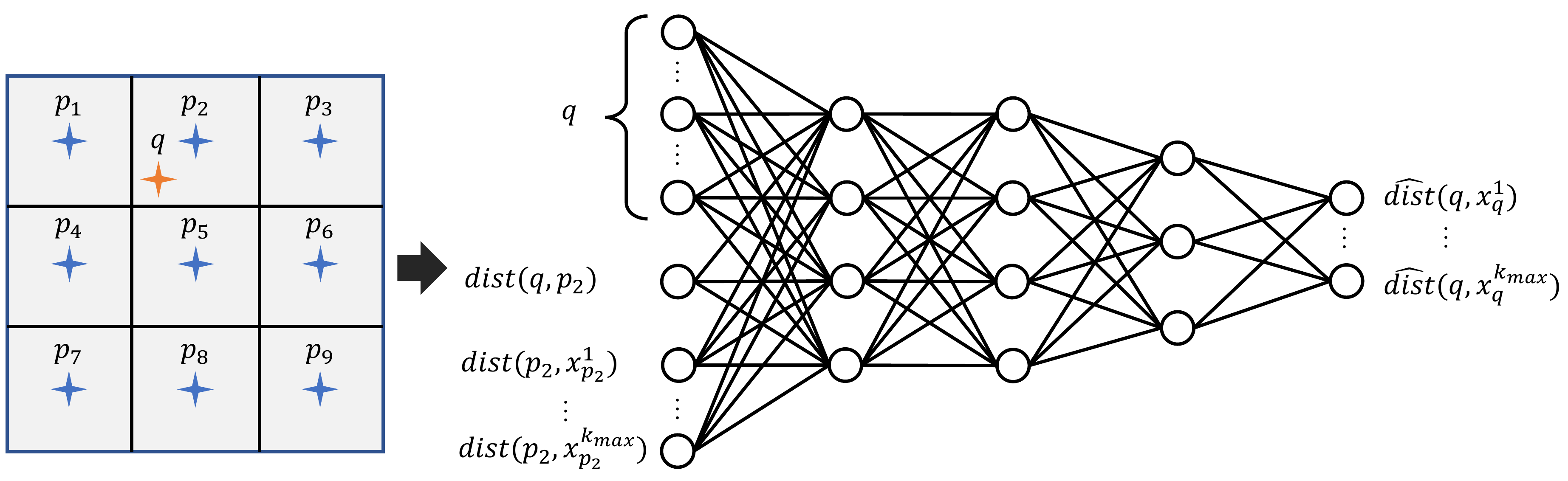}
	    \vspace{-3.0mm}
        \caption{The left part shows the data space and a grid.
        Blue stars centered at cells represent pivots and the red star represents a query $q$.
        The right part shows our five-layer fully connected neural network model.
        The input layer has the coordinates of $q$, $dist(q,p_{2})$, where $p_{2}$ is the nearest pivot of $q$, and $dist(p_{2},x_{p_{2}}^{k})$ ($k \in [1,k_{max}]$).
        It has three hidden layers, and its output layer is regarded as a vector $\langle \hat{dist}(q,x_{q}^{1}), ..., \hat{dist}(q,x_{q}^{k_{max}})\rangle$.}
        \label{figure_pivnet}
	\end{center}
\end{figure*}

\section{Our Solution}  \label{section_solution}
\noindent
\textbf{Main idea.}
The main requirements of the $k$-NN distance estimation problem are high efficiency and accuracy. 
To satisfy these requirements, our first idea uses pivot points.
Consider that a lot of pivots are distributed in the data space $\in \mathbb{R}^{d}$.
Specifically, the data space is divided equally by fine-grained grid cells and each cell has a pivot $p$ centered at the cell.
Given a query $q$, its corresponding cell can be found in $O(1)$ time.
Let $x_{p}^{i}$ be the $i$-th NN of $p$ in $X$.
From triangle inequality, if $x_{p}^{k} \notin \{x_{q}^{i} | i \in [1,k-1]\}$, we have
\begin{equation}
    dist(q,x_{q}^{k}) \leq dist(q,p) + dist(p,x_{p}^{k}).   \label{equation_triangle}
\end{equation}
Besides, in a very fine-grained grid, we can have $dist(q,p) \approx 0$.
This suggests that the distances to the $k$NNs of $q$ can be obtained from those of $p$.
Note that $dist(p,x_{p}^{i})$, where $i \in [1,k_{max}]$, can be obtained efficiently in a pre-processing phase (by using a tree index, e.g., $k$d-tree).
That is, for $k_{max} = O(1)$, this approach infers the $k$NN distances of $q$ in $O(1)$ time and its accuracy would be high.

Although the above idea functions well in theory, preparing such an unlimited number of pivots is space-consuming and is not feasible in practice.
Now the challenge is how to achieve the above high accuracy while keeping $O(1)$ inference time and the usefulness of pivots.
We overcome this challenge by using a neural network model, because it can effectively learn the distributions of real datasets.
We incorporate the idea that the $k$NN distances of a pivot near $q$ can be similar to those of $q$ into the neural network, to train an accurate estimation model.
This is our second idea.

\vsp
\noindent
\textbf{Modeling by a neural network.}
First, we divide each coordinate of the data space equally to build a grid offline.
The granularity is application-dependent and can be specified based on memory size.
Each cell of the grid has a pivot $p$ at its centroid.
Note that, given a query $q$, its nearest pivot is obtained in $O(1)$ time because pivots are centroids of their cells.

Then, to implement our main ideas, we consider the $k$NN distance estimation problem as a regression problem that outputs $\hat{dist}(q,x_{q}^{k})$.
By preparing a set $Q_{train}$ of training queries, we can obtain $dist(q,x_{q}^{k})$ offline.
Therefore, we can run supervised learning to train the regression model.
We employ a neural network to achieve this, motivated by the fact that it provides high inference accuracy in many problems, domains, and fields.
It is important to recall that the inference time has to be fast while minimizing estimation errors, as mentioned in Section \ref{section_introduction}.
To achieve this, a \textit{simple yet effective model} is required.
(Neural networks with complex and/or many layers \textit{cannot} achieve this, because they increase the number of running costly matrix multiplication.)
We specifically utilize a five-layer (three hidden layers) fully connected neural network and use ReLU \cite{nair2010rectified} as an activation function\footnote{Our experimental result shows that a deeper neural network does not yield a clear advantage.}.
As the input of this neural network, we use $q$, $k$, $dist(q,p)$, where $p$ is the nearest pivot of $q$, and $dist(p,x_{p}^{k})$.
Via stochastic gradient descent, this neural network is trained to minimize the following loss function:
\begin{equation*}
  \mathcal{L} = \frac{1}{k_{max}|Q_{train}|}\sum_{\langle q,k\rangle \in P}|dist(q,x_{q}^{k}) - \hat{dist}(q,x_{q}^{k})|,
\end{equation*}
where $P$ is a set of all pairs of $q \in Q_{train}$ and $k \in [1,k_{max}]$, and $\hat{dist}(q,x_{q}^{k})$ is obtained by the neural network $f$, i.e.,
\begin{equation*}
    \hat{dist}(q,x_{q}^{k}) = f(q,k,dist(q,p),dist(p,x_{p}^{k})).
\end{equation*}
By adding $dist(q,p)$ and $dist(p,x_{p}^{k})$ as feature values, its inference accuracy becomes higher than that of its variant without them while keeping simpleness, see Section \ref{section_experiment}.

\vsp
\noindent
\textbf{Optimization.}
The above neural network has several drawbacks.
First, it needs to deal with each $k \in [1,k_{max}]$ iteratively, as it outputs only $\hat{dist}(q,x_{q}^{k})$.
As described in Definition \ref{definition_problem}, when we need distances to $k$-NNs, this approach incurs $O(k)$ time.
For example, in Example \ref{example_density-estimation}, $k$ is not small, so this cost is not negligible.
Moreover, this approach incurs a long training time because we need to consider each pair of $q \in Q_{train}$ and $k \in [1,k_{max}]$.

To remove these drawbacks, we optimize our design of the neural network and propose PivNet, which is illustrated in Figure \ref{figure_pivnet}.
The last layer of this neural network has $k_{max}$ neurons, and this model is trained so that it can infer the distances to $k_{max}$-NNs \textit{at a time}.
This model, $f_{PivNet}$, requires $q$, $k$, $dist(q,p)$, where $p$ is the nearest pivot of $q$, and $\langle dist(p,x_{p}^{1}), ..., dist(p,x_{p}^{k_{max}})\rangle$ as input, and it outputs estimated $k_{max}$-NN distances.
That is,
\begin{align*}
  \hat{\mathbf{v}}_{q} &= f_{PivNet}(q, dist(q,p),\mathbf{v}_{p}), \text{where} \\
  \mathbf{v}_{p} &= \langle dist(p,x_{p}^{1}), ..., dist(p,x_{p}^{k_{max}})\rangle, \text{and}\\
  \mathbf{v}_{q} &= \langle \hat{dist}(q,x_{q}^{1}), ..., \hat{dist}(q,x_{q}^{k_{max}})\rangle.
\end{align*}
The loss function of PivNet is also optimized to deal with the distances to $k_{max}$-NNs at the same time while achieving accurate regression.
Let $||\cdot||_{1}$ be the $L_{1}$ norm of a given vector.
To obtain a similar accuracy to that of $f$, PivNet specifically utilizes the following loss function:
\begin{equation*}
  \mathcal{L}_{opt} = \frac{1}{|Q_{train}|}\sum_{q \in Q_{train}}\frac{||\mathbf{v}_{q} - \hat{\mathbf{v}}_{q}||_{1}}{k_{max}}.
\end{equation*}
(Notice that $\mathbf{v}_{q}$ is the ground truth.)
We train PivNet so that $\mathcal{L}_{opt}$ is minimized.
Then, as long as $k \leq k_{max}$, PivNet infers the distances to the $k$-NNs of $q$ in $O(1)$ time (for fixed hyper-parameters) without losing accurate estimation ability.

\vsp
\noindent
\textbf{Preparing query training set.}
A straightforward approach to preparing queries for training is to randomly sample points from $X$.
In this case, however, the model overfits the distribution of $X$.
When queries are specified uniformly at random in $\mathbb{R}^{d}$ as in Example \ref{example_density-estimation}, the performance of this model would degrade, particularly when queries are specified at sparse spaces.
To avoid this drawback, we augment training queries by generating synthetic ones that follow a uniform distribution in $\mathbb{R}^{d}$ (like pivots), and they are added into $Q_{train}$.
After that, for each $q \in Q_{train}$, we compute its $k$-NNs among $X\backslash Q_{train}$ and train PivNet.

\section{Experiment}    \label{section_experiment}
This section presents our experimental results.
All experiments were conducted on a Ubuntu 18.04 LTS machine with 3.0GHz Core i9-10980XE CPU and 128GB RAM.

\subsection{Setting}
\noindent
\textbf{Methods.}
We compared PivNet with the following approaches:
\begin{itemize}
    \setlength{\leftskip}{-4.0mm}
    \item   $k$d-tree \cite{bentley1975multidimensional}:
            This returns the exact answer by retrieving the $k$-NNs via a branch-and-bound algorithm.
    \item   LightGBM \cite{ke2017lightgbm}:
            This is a state-of-the-art gradient boosting decision tree-based regression model.
            For its input, we used the same feature values as those of PivNet, to be fair.
    \item   Pivot:
            This algorithm uses the right side of Equation (\ref{equation_triangle}) as inference of the $k$-NN distances.
    \item   QueryNet:
            This is a variant of PivNet and does not utilize pivots.
            This model also estimates the distances to $k_{max}$-NNs at a time.
    \item   PivNet-itr:
            This is a variant of PivNet, and it infers the distance to $k$-NN for each $k \in [1,k_{max}]$ iteratively.
\end{itemize}
The above methods were implemented by Python, were single threaded, and ran in-memory.
(We have no other competitors, because our problem has no prior works.)
QueryNet, PivNet-itr, and PivNet were trained in PyTorch.
For $k$d-tree, we used SciPy\footnote{\url{https://docs.scipy.org/doc/scipy/reference/generated/scipy.spatial.KDTree.html}}.

\vsp
\noindent
\textbf{Datasets.}
We used the following publicly available real datasets for evaluation.
In them, we removed points with missing values.
\begin{itemize}
    \setlength{\leftskip}{-4.0mm}
    \item   \textit{Crime} \cite{chan2020quad}:
            A set of 259,809 geo-location (i.e., 2D) points where crimes occurred in Atlanta, U.S.A.
    \item   \textit{HEPMASS} \cite{baldi2016parameterized}:
            A set of 10,049,359 5D mass feature datasets.
    \item   \textit{Household}\footnote{\url{https://archive.ics.uci.edu/ml/datasets/individual+household+electric+power+consumption}}:
            A set of 1,771,612 power consumption data measured in a house.
            We used 4 attributes (active power, reactive power, voltage, and intensity).
    \item   \textit{PAMAP2} \cite{reiss2012introducing}:
            A set of 2,643,873 5D-sensor data.
    \item   \textit{Places}\footnote{https://archive.org/details/2011-08-SimpleGeo-CC0-Public-Spaces}:
            A set of 9,033,486 geo-locations of public places in the U.S.A.
    \item   \textit{Wisdom} \cite{weiss2019smartphone}:
            A set of 4,476,481 3D gyroscopes sensors obtained from smartphones and smartwatches.
\end{itemize}

For each dataset, we randomly choose 100,000 points in the dataset and added them into $Q_{train}$.
Moreover, we added 100,000 points augmented by the approach presented in Section \ref{section_solution} into $Q_{train}$.
We used random 80\% points in $Q_{train}$ for training and the remaining points for validation.
We randomly choose another 10,000 points in the dataset, augmented 10,000 points in the above manner, and used them as test queries.
A set of the remaining points in the dataset was used as $X$.

\vsp
\noindent
\textbf{Hyper-parameters.}
We set $k_{max} = 50$.
For the grid introduced in Section \ref{section_solution}, we defined a constant $c$ to equally divide the coordinate of each domain.
For 2D and 3D datasets, $c = 2048$ and $c = 256$, respectively.
For 4D and 5D datasets, $c = 32$.

The numbers of neurons in the second and third layers were 64 on Crime and Places and 128 on the others.
The number of neurons in the fourth layer was 32 on all datasets.
The batch size was 500.
The learning rate of QueryNet was 0.1, 0.1, 0.1, 0.2, 0.2, and 0.1 on Crime, HEPMASS, Household, PAMAP2, Places, and Wisdom, respectively.
Also, that of PivNet(-itr) was 0.03 (0.01), 0.02 (0.003), 0.01 (0.005), 0.04 (0.003), 0.01 (0.003), and 0.02 (0.005) on Crime, HEPMASS, Household, PAMAP2, Places, and Wisdom, respectively.

\vsp
\noindent
\textbf{Evaluation criteria.}
To measure the accuracy of each inference model, we computed the average and median of $MAE_{q}$, which is defined in Equation (\ref{equation_mae}), and those of $MAPE_{q}$ among all queries in the test query set.
Note that $MAPE_{q}$ is obtained as follows:
\begin{equation*}
    MAPE_{q} = \frac{1}{k_{max}}\sum_{k = 1}^{k_{max}}\frac{|dist(q,x_{q}^{k}) - \hat{dist}(q,x_{q}^{k})|}{dist(q,x_{q}^{k})}.
\end{equation*}
As for running time, we measured the average time to obtain the $k_{max}$-NN distances of a given query.
In addition, we measured the training time of each ML model.

\subsection{Result} \label{section_experiment-result}
Tables \ref{table_error} and \ref{table_error_} respectively show the results of MAE and MAPE (smaller is better), while Table \ref{table_time} shows the result of time.

\begin{table*}[!t]
    \begin{center}
        \caption{Average and median of MAE}
        \label{table_error}
        \vspace{-4.0mm}
        \begin{tabular}{l|l|cccccc} \toprule
                                        &           & Crime     & HEPMASS   & Household & PAMAP2    & Places    & Wisdom    \\ \midrule
            \multirow{2}{*}{LightGBM}   & Average   & 0.00057   & 0.05989   & 0.09785   & 0.09005   & 0.09216   & 0.03367   \\ 
                                        & Median    & 0.00014   & 0.01714   & 0.02031   & 0.04586   & 0.00284   & 0.01168   \\ \hline                            
            \multirow{2}{*}{Pivot}      & Average   & 0.00049   & 0.12295   & 0.28569   & 0.38003   & 0.00811   & 0.05177   \\ 
                                        & Median    & 0.00046   & 0.12486   & 0.28337   & 0.37429   & 0.00742   & 0.05077   \\ \hline
            \multirow{2}{*}{QueryNet}   & Average   & 0.00075   & 0.02225   & 0.02949   & 0.07335   & 0.07702   & 0.04213   \\
                                        & Median    & 0.00048   & 0.01597   & 0.01691   & 0.04512   & 0.02336   & 0.02483   \\ \hline
            \multirow{2}{*}{PivNet-itr} & Average   & 0.00018   & 0.00977   & 0.01024   & 0.03828   & 0.00711   & 0.01184   \\
                                        & Median    & 0.00012   & 0.00578   & 0.00536   & 0.02499   & 0.00313   & 0.00806   \\ \hline
            \multirow{2}{*}{PivNet}     & Average   & 0.00019   & 0.01965   & 0.01852   & 0.04768   & 0.01113   & 0.01038   \\
                                        & Median    & 0.00013   & 0.01117   & 0.00916   & 0.03528   & 0.00320   & 0.00748   \\ \bottomrule
        \end{tabular}
        \vspace{-3.0mm}
    \end{center}
\end{table*}
\begin{table*}[!t]
    \begin{center}
        \caption{Average and median of MAPE}
        \label{table_error_}
        \vspace{-4.0mm}
        \begin{tabular}{l|l|cccccc} \toprule
                                        &           & Crime & HEPMASS   & Household & PAMAP2    & Places    & Wisdom    \\ \midrule
            \multirow{2}{*}{LightGBM}   & Average   & 0.14  & 0.14      & 0.33      & 0.38      & 0.78      & 0.37      \\ 
                                        & Median    & 0.07  & 0.06      & 0.14      & 0.06      & 0.11      & 0.05      \\ \hline                            
            \multirow{2}{*}{Pivot}      & Average   & 0.35  & 0.95      & 7.32      & 2.16      & 4.19      & 0.93      \\ 
                                        & Median    & 0.15  & 0.40      & 1.46      & 0.45      & 0.26      & 0.19      \\ \hline
            \multirow{2}{*}{QueryNet}   & Average   & 0.38  & 0.10      & 0.19      & 0.12      & 6.45      & 0.34      \\
                                        & Median    & 0.27  & 0.06      & 0.08      & 0.07      & 0.74      & 0.12      \\ \hline
            \multirow{2}{*}{PivNet-itr} & Average   & 0.14  & 0.04      & 0.11      & 0.07      & 1.09      & 0.12      \\
                                        & Median    & 0.06  & 0.02      & 0.03      & 0.04      & 0.13      & 0.04      \\ \hline
            \multirow{2}{*}{PivNet}     & Average   & 0.14  & 0.07      & 0.14      & 0.17      & 0.89      & 0.11      \\
                                        & Median    & 0.06  & 0.04      & 0.04      & 0.05      & 0.14      & 0.03      \\ \bottomrule
        \end{tabular}
        \vspace{-3.0mm}
    \end{center}
\end{table*}

\begin{table*}[!t]
    \begin{center}
        \caption{Average time to obtain $k_{max}$-NN distances and training time}
        \label{table_time}
        \vspace{-4.0mm}
        \begin{tabular}{l|l|cccccc} \toprule
                                        &                           & Crime     & HEPMASS   & Household & PAMAP2    & Places    & Wisdom    \\ \midrule
            $k$d-tree                   & Retrieval time [microsec] & 56.04     & 148.02    & 104.68    & 154.18    & 54.46     & 65.77     \\ \hline
            \multirow{2}{*}{LightGBM}   & Inference time [microsec] & 13.20     & 11.35     & 11.44     & 11.36     & 13.67     & 13.47     \\
                                        & Training time [min]       & 0.32      & 0.34      & 0.32      & 0.34      & 0.32      & 0.31      \\ \hline
            \multirow{1}{*}{Pivot}      & Inference time [microsec] & 1.36      & 1.60      & 1.15      & 1.48      & 1.62      & 1.67      \\ \hline
            \multirow{2}{*}{QueryNet}   & Inference time [microsec] & 1.50      & 2.02      & 2.23      & 2.63      & 1.05      & 2.14      \\
                                        & Training time [min]       & 6         & 2         & 12        & 5         & 7         & 7         \\ \hline
            \multirow{2}{*}{PivNet-itr} & Inference time [microsec] & 191.49    & 193.84    & 192.37    & 193.37    & 192.38    & 196.04    \\
                                        & Training time [min]       & 119       & 575       & 217       & 746       & 47        & 232       \\ \hline
            \multirow{2}{*}{PivNet}     & Inference time [microsec] & 3.24      & 3.94      & 5.65      & 5.13      & 3.58      & 4.81      \\
                                        & Training time [min]       & 15        & 6         & 9         & 3         & 10        & 38        \\ \bottomrule
        \end{tabular}
    \end{center}
\end{table*}

\vsp
\noindent
\textbf{Effectiveness of using neural network.}
First, let us compare PivNet (or PivNet-itr) with Pivot to study the effectiveness of utilizing neural network.
Tables \ref{table_error} and \ref{table_error_} show that the accuracy of PivNet(-itr) is better than that of Pivot.
This result suggests that our neural network does not simply use the $k$-NN distances of pivots as inference but learns an accurate regression model.
Notice that Pivot provides upper-bounds of $k$-NN distances, see Equation (\ref{equation_triangle}).
These bounds become loose as $d$ increases.
For example, Pivot yields a small MAE on Crime and Places ($d = 2$) but its accuracy degrades on PAMAP2 ($d = 5$).

Moreover, we see that PivNet yields better accuracy than LightGBM in most tests.
This result also justifies our choice of utilizing a neural network.

\begin{table*}[!t]
    \begin{center}
        \caption{MAE of PivNet with different $k$}
        \label{table_error-k}
        \vspace{-4.0mm}
        \begin{tabular}{lcccccc} \toprule
                            & Crime     & HEMPASS   & Household & PAMAP2    & Places    & Wisdom    \\ \midrule
            $k \in [1,10]$  & 0.00023   & 0.02214   & 0.02001   & 0.05660   & 0.02167   & 0.01315   \\ 
            $k \in [11,20]$ & 0.00019   & 0.01930   & 0.01790   & 0.04582   & 0.00995   & 0.01029   \\
            $k \in [21,30]$ & 0.00019   & 0.01888   & 0.01775   & 0.04497   & 0.00648   & 0.00961   \\
            $k \in [31,40]$ & 0.00017   & 0.01888   & 0.01827   & 0.04525   & 0.00733   & 0.00958   \\
            $k \in [41,50]$ & 0.00019   & 0.01926   & 0.01890   & 0.04665   & 0.01097   & 0.00951   \\ \bottomrule
        \end{tabular}
        \vspace{-3.0mm}
    \end{center}
\end{table*}

\vsp
\noindent
\textbf{Effectiveness of using pivots.}
We next investigate the usefulness of pivots by comparing PivNet with QueryNet.
From Table \ref{table_error}, we see that PivNet is always better than QueryNet.
Also, Table \ref{table_error_} shows that PivNet is better than QueryNet with only a single exception of the average case of PAMAP2.
(Recall that, as mentioned in Section \ref{section_preliminary}, we weight MAE.)
For example, PivNet provides about 3.9x (3.7x), 1.1x (1.4x), 1.9x (1.8x), 1,5x (1.3x), 6.9x (7.3x), and 4.1x (3.3x) less MAE than QueryNet in the average (median) case on Crime, HEPMASS, Household, PAMAP2, Places, and Wisdom, respectively.
Clearly, pivots and their $k$-NN distances support accurate model training.

\vsp
\noindent
\textbf{Comparison with PivNet-itr.}
For PivNet and PivNet-itr, we study how the difference in neural network structures impacts their performances.
Tables \ref{table_error} and \ref{table_error_} show that their accuracy is generally competitive.
This actually suggests that, although PivNet learns every $k \in [1,k_{max}]$ at the same time, it still effectively learns an accurate regression model.
When PivNet-itr yields a less error (e.g., HEPMASS case) and applications need a single $k$, PivNet-itr can be an option.

\vsp
\noindent
\textbf{Inference time.}
We turn our attention to time to obtain $k$-NN distances.
From Table \ref{table_time}, we first observe that Pivot is the fastest.
This is reasonable, because it simply computes Equation (\ref{equation_triangle}), i.e., just ``$+$'' operation is required after computing $dist(p,q)$.
Recall that the accuracy of Pivot is the worst in most cases.

Second, QueryNet is faster than PivNet(-itr).
This is also a natural observation.
QueryNet does not use pivots, so the number of neurons in its input layer is smaller and it does not need to compute the cell of a given query and the distance between the query and its nearest pivot.
These provide the time difference.
We next see that PivNet-itr is the slowest, i.e., it is outperformed by the exact approach $k$d-tree.
Since PivNet-itr needs $k_{max}$ times inferences, this is also a reasonable result.
(However, if we need a single $k$, its inference time becomes $k_{max}$ times faster.)

PivNet is slower than Pivot and QueryNet, but its accuracy is much better than those of them and its inference is absolutely fast and stable (as it needs a constant time for inference).
In addition, its inference times are about 17.3x (4.1x), 37.6x (2.9x), 18.5x (2.0x), 39.8x (2.9x), 15.2x (3.8x), and 13.7x (2.8x) faster than those of $k$d-tree (LightGBM) on Crime, HEPMASS,  Household, PAMAP2, Places, and Wisdom, respectively.
The above results demonstrate that PivNet clearly has the best trade-off relationship between inference time and accuracy among the evaluated methods. 

\vsp
\noindent
\textbf{Training times} of ML methods are also shown in Table \ref{table_time}.
LightGBM has less training time than the neural network models, but PivNet also spent reasonable time for \textit{one-time} training.
For neural network models, their tendencies of training time clearly follow that of inference time, because the training time also depends on the structure of a given neural network.
Although we assume static datasets, PivNet can deal with dataset updates if the update frequency is moderate (e.g., updates are reflected in a batch manner once a few hours), since its training time is not significant.

\begin{figure*}[!t]
	\begin{center}
	    \subfigure[Exact on Crime]{%
		    \includegraphics[width=0.24\linewidth]{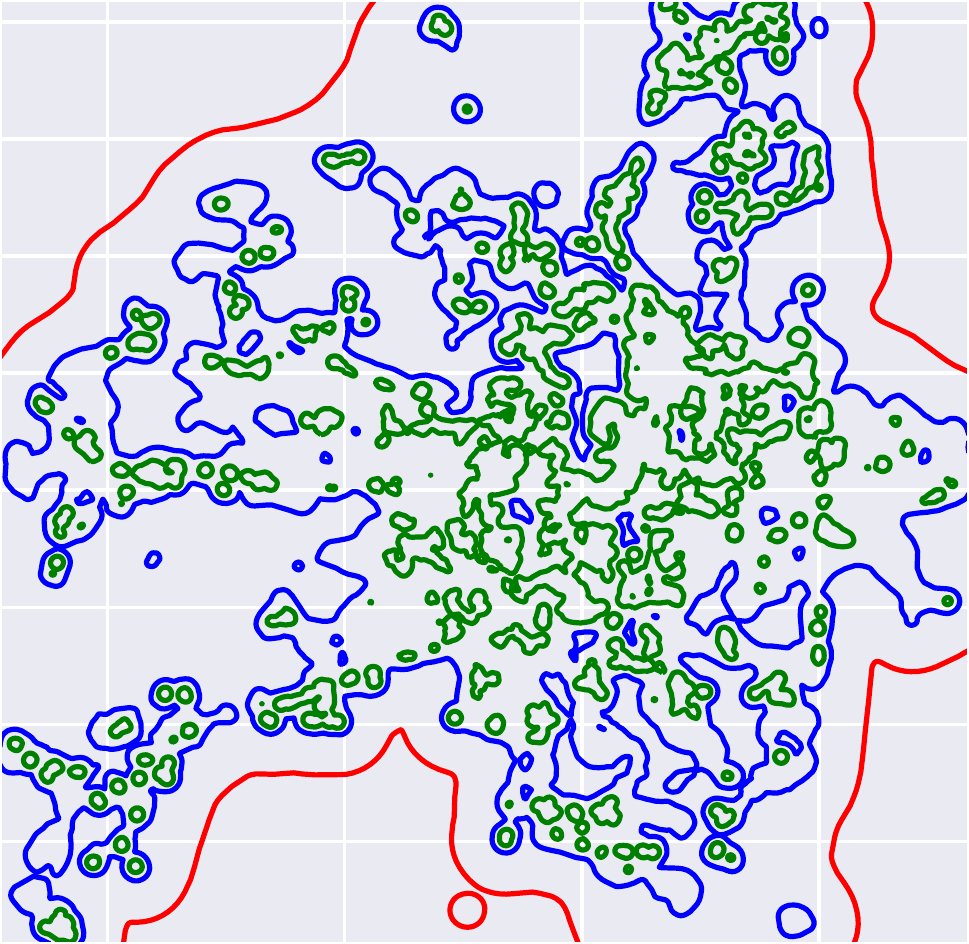}	    \label{fig_knndv-crime}}
        \subfigure[Estimation by PivNet on Crime]{%
		    \includegraphics[width=0.24\linewidth]{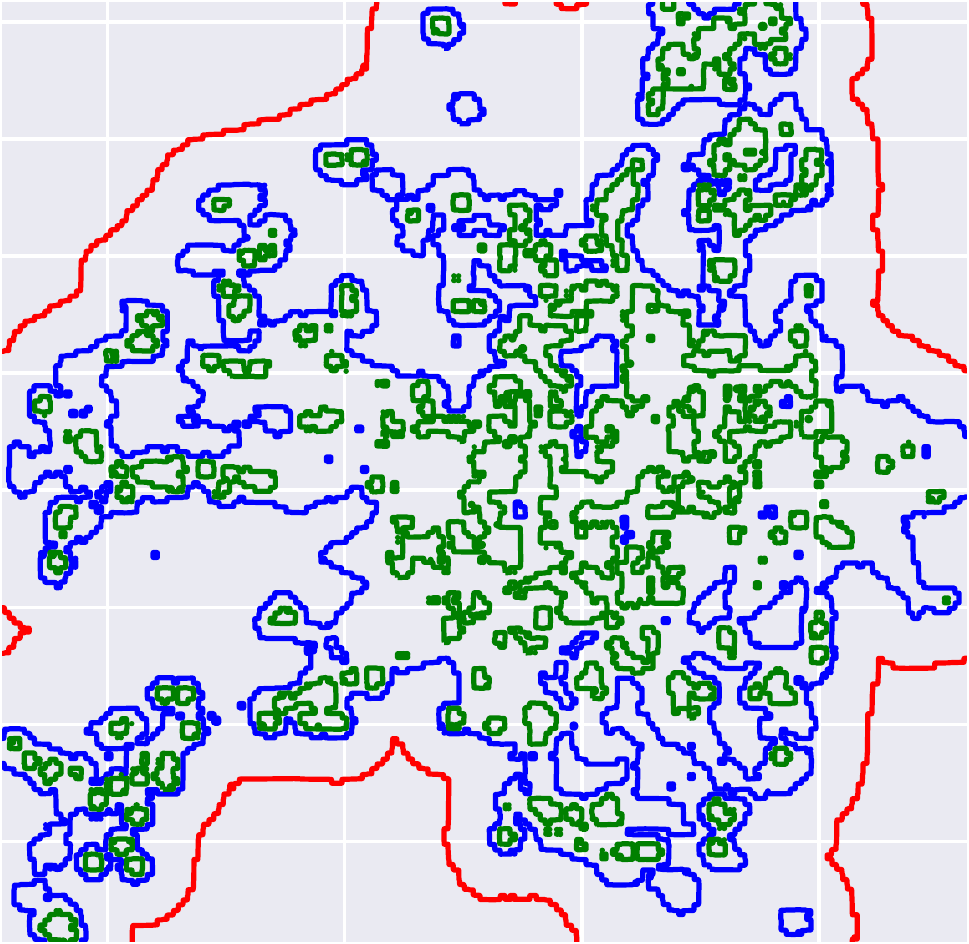}	\label{fig_knndv-crime-est}}
	    \subfigure[Exact on Places]{%
		    \includegraphics[width=0.24\linewidth]{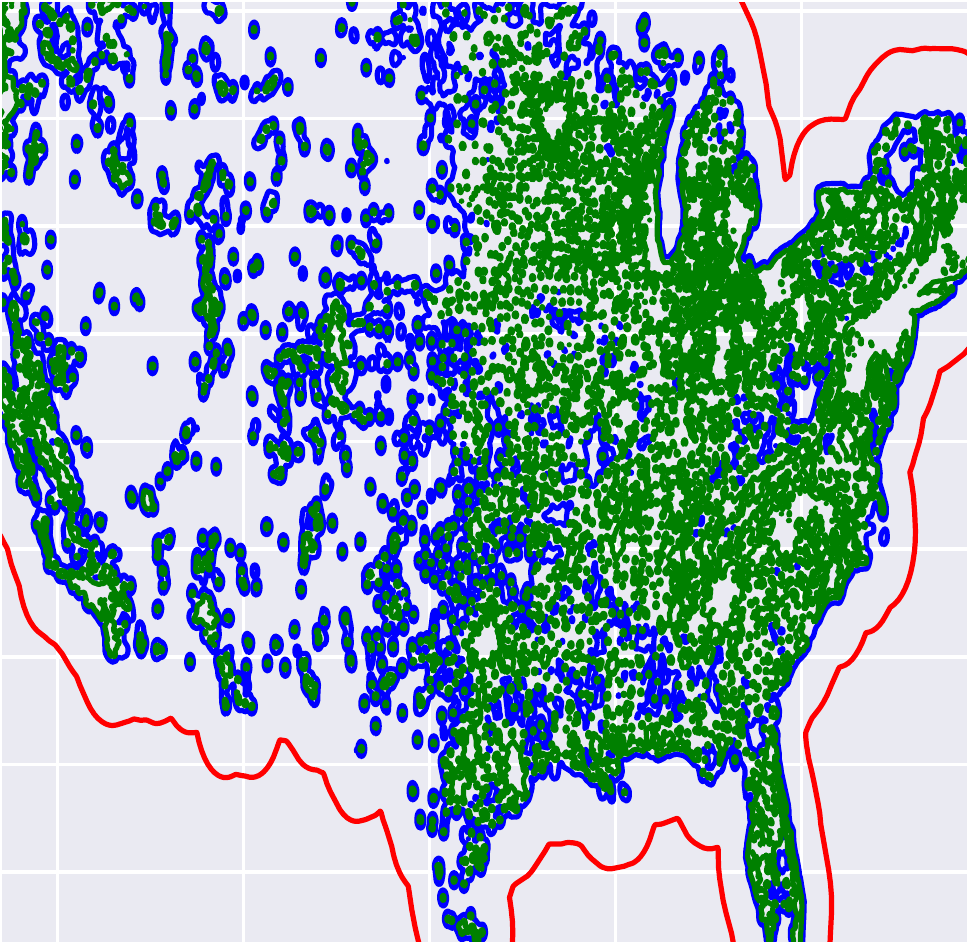}	    \label{fig_knndv-places}}
        \subfigure[Estimation by PivNet on Places]{%
		    \includegraphics[width=0.24\linewidth]{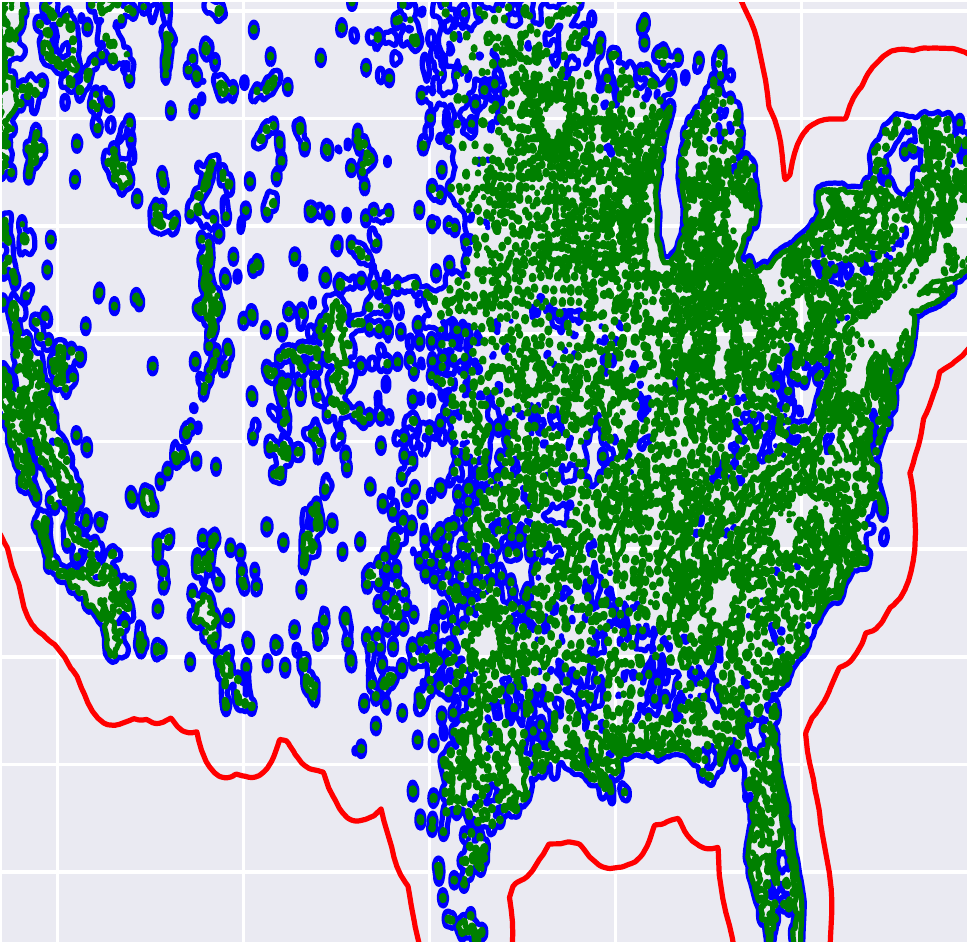}	\label{fig_knndv-places-est}}
		\vspace{-4.0mm}
        \caption{$k$-NN density visualization}
        \label{figure_knndv}
        \vspace{-4.0mm}
	\end{center}
\end{figure*}

\vsp
\noindent
\textbf{Impact of $k$.}
We next study the error of PivNet by decomposing $[1,50]$ into five disjoint subsets as shown in Table \ref{table_error-k}.
(We omit MAPE, since its tendency simply follows that of MAE.)
We observe that PivNet has a higher MAE in the case of $k \in [1,10]$ compared with the other cases.
When $k$ is small, a density difference in $X$ clearly appears.
As the number of points existing in sparse spaces is much smaller than that of points in dense spaces, perhaps neural networks still face difficulty of estimating accurate $k$-NN distances for such sparse points.
However, the errors are still sufficiently small.
(Recall that the inference time of PivNet is not affected by $k$ iff $k \leq k_{max}$.)

\vsp
\noindent
\textbf{Is a deeper neural network better?}
Last, we address this question by using PivNet-5, which has five hidden layers (recall that PivNet originally has three hidden layers).
The MAE (average case) of PivNet-5 was 0.00019, 0.02399, 0.02051, 0.04859, 0.01163, and 0.01166 on Crime, HEPMASS, Household, PAMAP2, Places, and Wisdom, respectively.
Also, its inference time [microsec] was 3.62, 5.19, 5.00, 5.13, 5.74, and 5.37 on Crime, HEPMASS, Household, PAMAP2, Places, and Wisdom, respectively.
Comparing with the results in Tables \ref{table_error} and \ref{table_time}, PivNet-5 does not improve accuracy and its inference time is slower.
We therefore conclude that \textbf{deeper neural networks are not necessary for PivNet and it can learn an accurate model with a small number of hidden layers}.

\begin{table*}[!t]
    \begin{center}
        \caption{Precision and recall in the problems of distance-based outlier detection}
        \label{table_dod-accuracy}
        \vspace{-3.0mm}
        \begin{tabular}{ccccccccc} \toprule
                        & \multicolumn{2}{c}{HEPMASS}   & \multicolumn{2}{c}{Household} & \multicolumn{2}{c}{PAMAP2}    & \multicolumn{2}{c}{Wisdom}    \\ \cmidrule{2-9}
                        & Precision & Recall            & Precision & Recall            & Precision & Recall            & Precision & Recall            \\ \cmidrule{2-9}
            $(r,k)$-DOD & 0.92      & 0.94              & 0.85      & 0.94              & 0.86      & 0.86              & 0.94      & 0.95              \\ 
            $(N,k)$-DOD & -         & 0.93              & -         & 0.89              & -         & 0.86              & -         & 0.95              \\ \bottomrule
        \end{tabular}
        \vspace{-4.0mm}
    \end{center}
\end{table*}

\section{Case Study}    \label{section_case-study}
We tested our $k$-NN distance estimation model in the application examples in Section \ref{section_introduction}.
In this section, we set $k_{max} = 100$, and the hyper-parameters were the same as those in Section \ref{section_experiment}.
We computed the exact results by using a $k$d-tree.

\subsection{$k$-NN Density Visualization}
\textbf{Setting.}
We used Crime and Places in this case study.
For each dataset, we made $1,000 \times 1,000$ pixels by equally dividing each domain.
We used the centroid of each pixel as a query and computed its $k$-NN density from Equation (\ref{equation_knnd}), where $k = 100$.
The result is visualized by a contour diagram, and its interval is determined by using 20\%, 60\%, and 90\% percentiles.

\vsp
\noindent
\textbf{Result.}
Figure \ref{figure_knndv} illustrates the visualization results on Crime and Places.
Note that green, blue, and red show high, middle, and low densities, respectively.
For example, the number of crimes is higher in the green areas of Crime.
Although there are minor differences, PivNet displays almost the same result as the exact solution.
Therefore, PivNet would not hinder the detection of information from the visualization results.

PivNet's ($k$d-tree's) time to compute (estimate) the $k$-NN density of all pixels was 3.1 (16.2) and 3.1 (35.8) seconds on Crime and Places, respectively.
PivNet is much faster than $k$d-tree and completes the computation within a few seconds for a million pixels.

\subsection{Distance-based Outlier Detection}   \label{section_dod}
We next consider the distance-based outlier detection problems defined below.

\begin{defi}[\textsc{$(r,k)$-distance-based outlier detection problem} \cite{knox1998algorithms}]   \label{definition_dod}
Given a set $X$ of points, $k$, and a distance-threshold $r$, this problem is to detect all points $x \in X$ such that the number of points $x'$ satisfying $dist(x,x') \leq r$ is less than $k$.
\end{defi}

\begin{defi}[\textsc{$(N,k)$-distance-based outlier detection problem} \cite{ramaswamy2000efficient}]
Given a set $X$ of points, $k$, and $N$, this problem finds $N$ points $x \in X$ whose distances to their $k$-th nearest neighbor are the largest among the points in $X$.
\end{defi}

\noindent
Both problems can be solved by evaluating the distance to the $k$-th nearest neighbor for each $x \in X$.

\vsp
\noindent
\textbf{Setting.}
In this study, we set $k = 50$.
For the $(r,k)$-distance-based outlier detection ($(r,k)$-DOD) problem, we set $r$ so that the number of outliers is 1,000.
Similarly, we set $N = 1,000$ for the $(N,k)$-DOD problem.
To measure the accuracy, we used \textit{precision} and \textit{recall}.
Let $TP$, $FP$, and $FN$ be the numbers of true positives, false positives, and false negatives, respectively.
Precision and recall are defined as follows:
\begin{equation*}
    Precision = \frac{TP}{TP + FP}, \;Recall = \frac{TP}{TP + FN}.
\end{equation*}
For evaluation, we used PivNet-itr (PivNet) on HEPMASS, Household, PAMAP2, and Wisdom.

\begin{table}[!t]
    \begin{center}
        \caption{Running time [sec] for distance-based outlier detection}
        \label{table_dod-time}
        \vspace{-4.0mm}
        \begin{tabular}{ccccc} \toprule
                        & HEPMASS   & Household & PAMAP2    & Wisdom    \\ \midrule
            $k$d-tree   & 964       & 93        & 248       & 102       \\ 
            Ours        & 21        & 3         & 7         & 13        \\ \bottomrule
        \end{tabular}
        \vspace{-4.0mm}
    \end{center}
\end{table}

\vsp
\noindent
\textbf{Result.}
Table \ref{table_dod-accuracy} shows the accuracy of our estimation-based detection.
We see that its detection accuracy is high (i.e., both $FP$ and $FN$ are small).
In particular, we achieve almost perfect recall on HEPMASS, Household, and Wisdom.
This result also confirms its high inference accuracy.

The detection time is shown in Table \ref{table_dod-time}.
Our approach is again much faster than $k$d-tree.
The time complexity of $k$d-tree for retrieving $k$-NNs is $O(kn^{1-1/d})$, where $n = |X|$.
It hence incurs $O(kn^{2-1/d})$ time for the distance-based outlier detection problems\footnote{This is faster than other works \cite{angiulli2009dolphin, bay2003mining, orair2010distance, tao2006mining}, because they incur $O(n^2)$ time.}, and it takes a long time for DOD on large datasets, e.g., HEPMASS.
On the other hand, for each point in $X$, we need only $O(1)$ time to evaluate (estimate) the $k$NN distance, thus our approach needs $O(n)$ time for the problems.
Its scalability to large datasets is hence obvious.
PivNet(-itr) can be a good option to quickly deal with such datasets of applications that allow approximate results.
This is often the case, because spatial data analysis requires interaction \cite{dong2020marviq, wang2015spatial}, e.g., outliers are evaluated while varying $k$, $r$, or $N$.

\begin{figure*}[!t]
	\begin{center}
	    \subfigure[Visualization of Syn]{%
		    \includegraphics[width=0.23\linewidth]{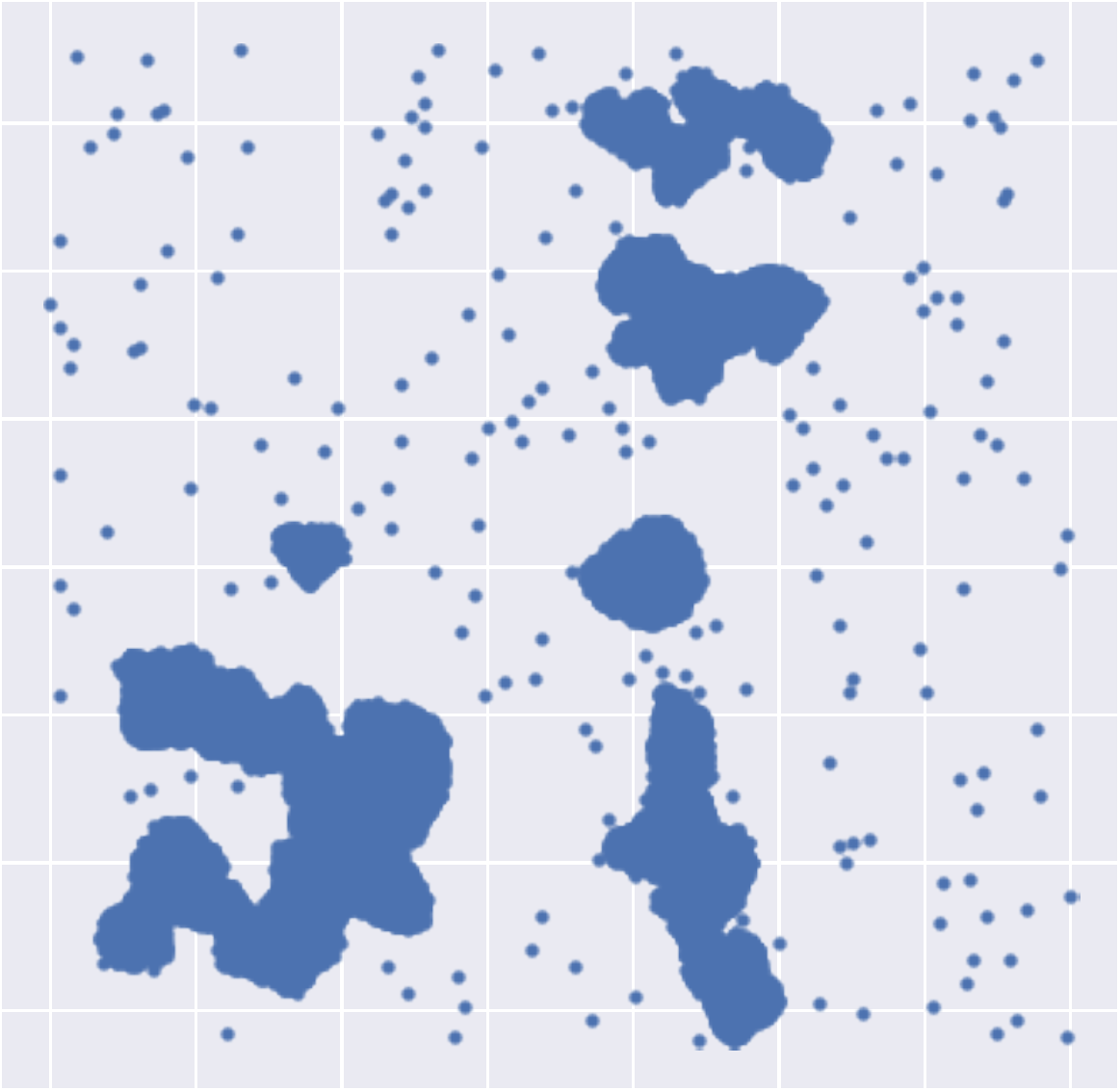}	    \label{fig_syn}}
	    \subfigure[Clustering result]{%
		    \includegraphics[width=0.23\linewidth]{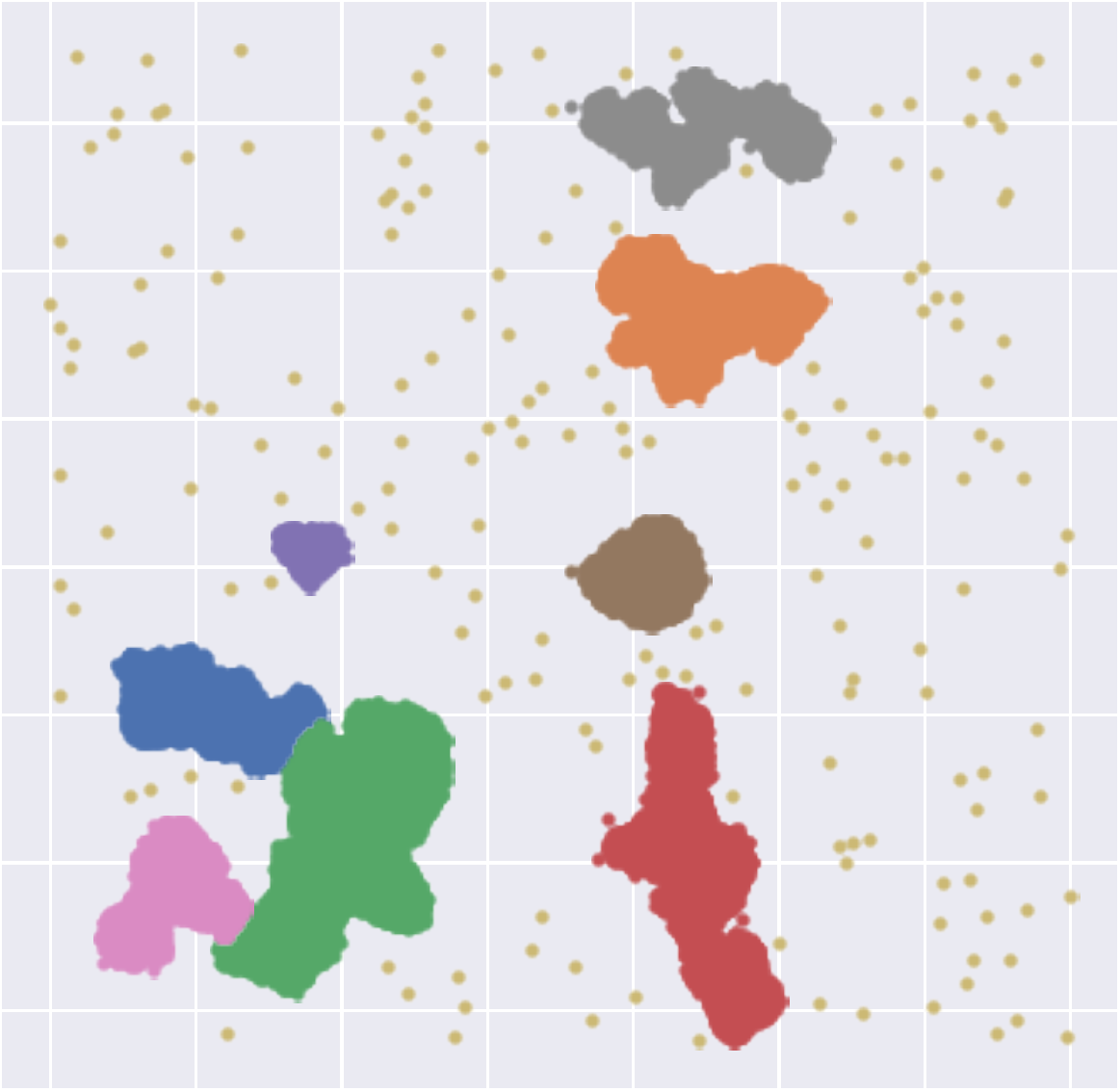}	    \label{fig_cluster}}
        \subfigure[Clustering result via estimated $d_{cut}$]{%
		    \includegraphics[width=0.23\linewidth]{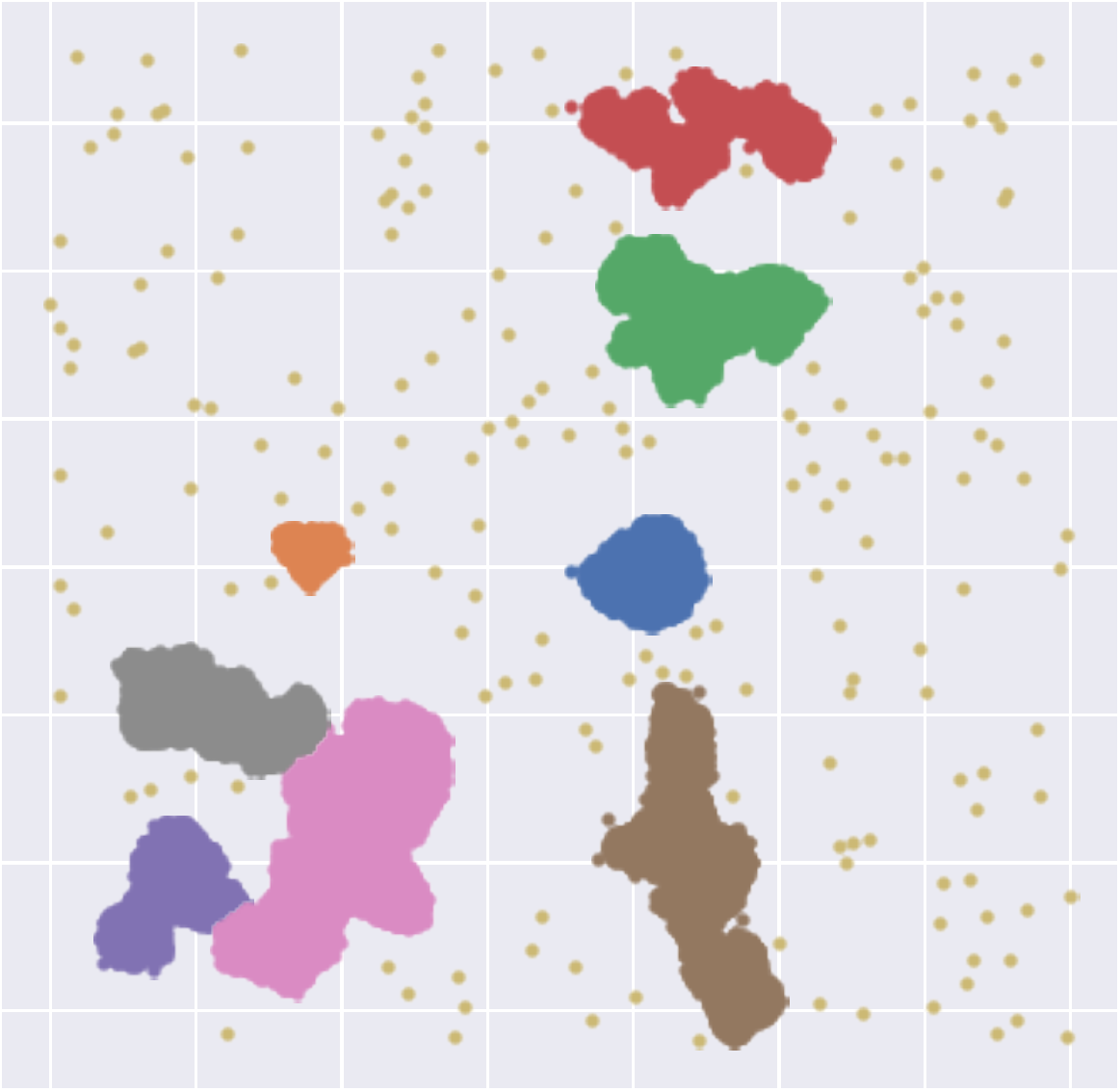}	\label{fig_cluster-est}}
		    
	    \subfigure[Decision graph]{%
		    \includegraphics[width=0.25\linewidth]{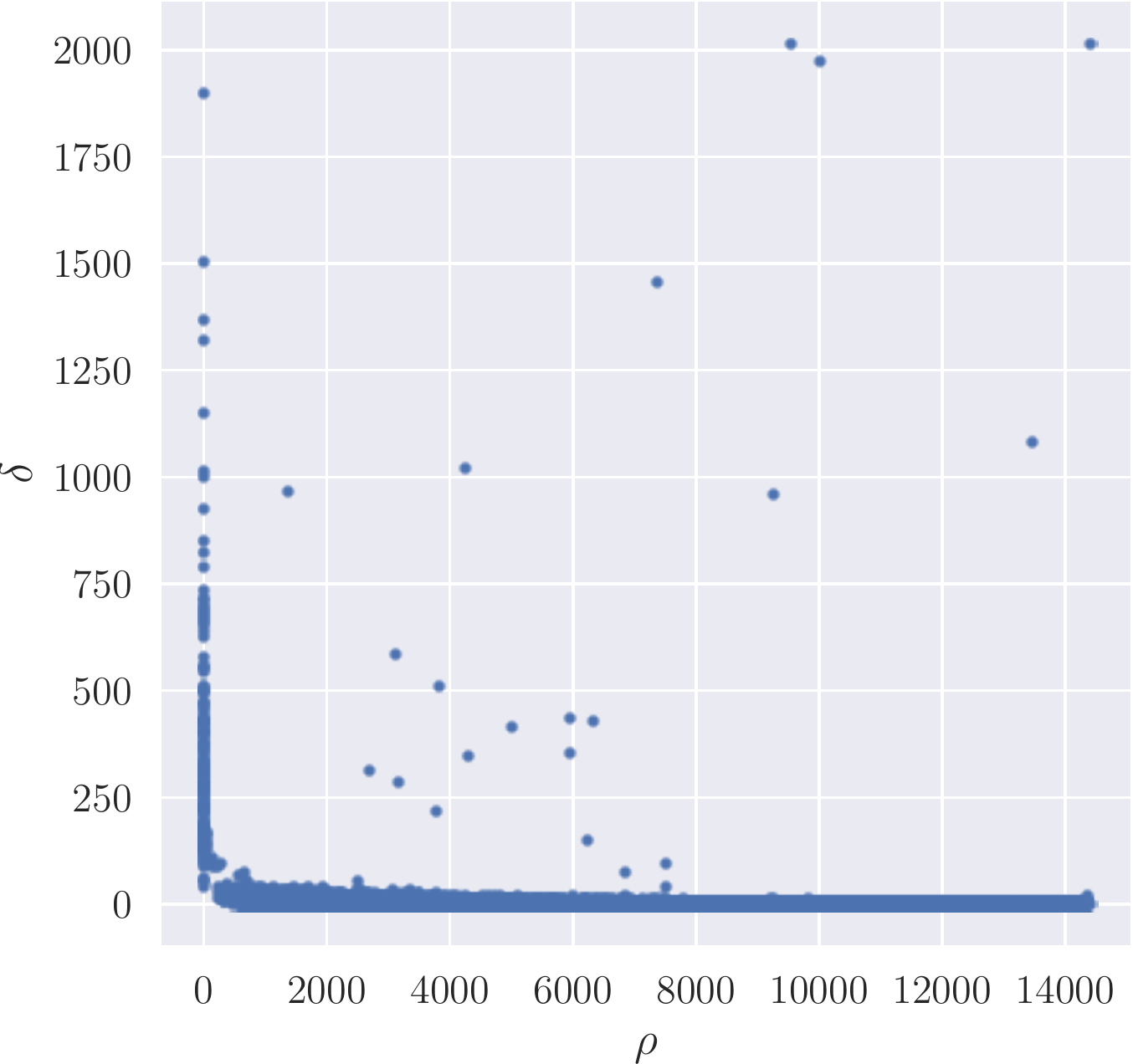}	    \label{fig_decision-graph}}
        \subfigure[Decision graph via estimated $d_{cut}$]{%
		    \includegraphics[width=0.25\linewidth]{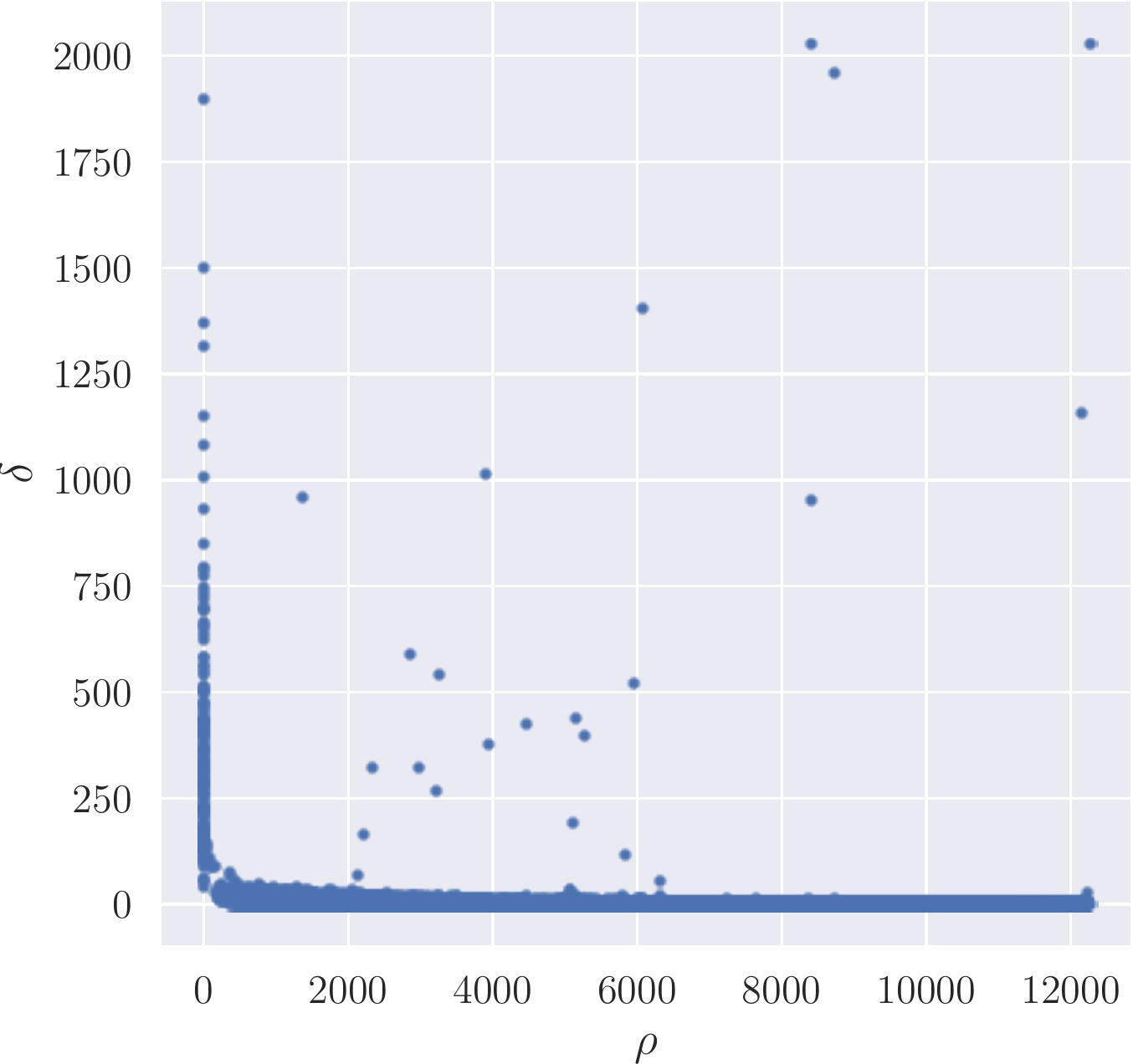}	\label{fig_decision-graph-est}}
		\vspace{-4.0mm}
        \caption{Comparison of clustering results and decision graphs via reverse engineering}
        \label{figure_decision-graph}
        \vspace{-3.0mm}
	\end{center}
\end{figure*}

\subsection{Reverse Engineering for Density-Peaks Clustering}   \label{section_case-study-dpc}
We next consider the estimation of a missing parameter of density-based clustering by utilizing PivNet.
As density-based clustering, we focus on density-peaks clustering (DPC) \cite{rodriguez2014clustering}.
This is a comparatively new algorithm but already has many applications \cite{amagata2021dpc}.
(For another famous density-based clustering DBSCAN, a reverse engineering-like tool has been already provided in \cite{campello2015hierarchical}.)

For each $x_{i} \in X$, DPC computes the local density $\rho_{i}$, which is the number of points $x_{j}$ such that $dist(x_{i},x_{j}) \leq d_{cut}$ , where $d_{cut}$ is a user-specified distance threshold.
In addition, for each $x_{i} \in X$, DPC computes its dependent point, which is the nearest point of $x_{i}$ with higher local density than $\rho_{i}$, and the dependent distance $\delta_{i}$, which is the distance between $x_{i}$ and its dependent point.
As with Definition \ref{definition_dod}, DPC removes noises that have less local density than $\rho_{min}$.
It is intuitively seen that, if a non-noise point $x$ has a large dependent distance, it is a density-peak in the space centered at $x$.
DPC employs this concept and regards it as a cluster center.
Consider the dataset illustrated in Figure \ref{fig_syn}.
After we compute $\rho_{i}$ and $\delta_{i}$ for each $x_{i} \in X$, $\langle \rho_{i},\delta_{i}\rangle$ can be mapped into a 2-dimensional space called a decision graph, see Figure \ref{figure_decision-graph}.
It visualizes that there are some points with large dependent distances, i.e., cluster centers.
By setting a threshold $\delta_{min}$ of dependent distance (e.g., $\delta_{min} = 800$ in Figure \ref{fig_decision-graph}), DPC determines cluster centers, and the other non-noise points belong to the same clusters as their dependent points.

It is often the case that analysis (e.g., clustering) results are known but (some of) their parameters are missing \cite{kalashnikov2018fastqre}.
In the context of DPC, it is possible that we have cluster labels and a decision graph but $d_{cut}$ is missing.
To reproduce the results, a straightforward approach is to run DPC while testing some values of $d_{cut}$.
However, DPC needs $o(n^2)$ time \cite{amagata2021dpc} (on low-dimensional datasets), so running DPC multiple times is not promising.
We alleviate this concern by estimating $d_{cut}$ and consider how to achieve this.

For simplicity, let us assume that $\rho_{min}$ and the number of noises (denoted by $m$) are known.
(They are obtained from cluster labels and the decision graph.)
From Section \ref{section_dod}, we can consider that each noise point $x$ has the top-$m$ largest distance between $x$ and its $\rho_{min}$-NN.
Therefore, by using the top-$m$ largest $\rho_{min}$-NN distance as $d_{cut}$, we can (approximately) distinguish non-noise points from noises and then reproduce the clustering result.

To test this idea, we used the synthetic dataset Syn in Figure \ref{fig_syn}, which was generated based on a random walk model \cite{gan2015dbscan}.
Syn has 200,000 points and eight density-peaks.
The exact clustering result with $d_{cut} = 200$ and its decision graph are respectively illustrated in Figures \ref{fig_cluster} and \ref{fig_decision-graph} (best viewed in color).
We set $\rho_{min} = 50$, and then Syn has 186 noises.
Next, we used PivNet to estimate the $\rho_{min}$-NN distance for each point.
As a result, the 186-th largest 50-NN distance was 179.23.
We used this value as the estimated $d_{cut}$ and ran DPC.
The clustering result and decision graph are depicted in Figures \ref{fig_cluster-est} and \ref{fig_decision-graph-est}, respectively.
It can be seen that our idea reproduces the clustering result almost perfectly.
This reproduction needed only one-time DPC and $O(n\log n)$ time to estimate $d_{cut}$ ($O(n)$ time for $\rho_{min}$-NN distance estimation and $O(n\log n)$ time to sort the distances), demonstrating that PivNet efficiently helps reverse engineering.

\begin{table*}[!t]
    \begin{center}
        \caption{Running time [microsec] for (approximate) $k$-NN search}   \label{table_aknn-time}
        \vspace{-3.0mm}
        \begin{tabular}{l|c|cccccc} \toprule
                                            & $k$   & Crime & HEPMASS   & Household & PAMAPA2   & Places    & Wisdom    \\ \midrule
            \multirow{4}{*}{Exact}          & 25    & 47    & 122       & 87        & 116       & 47        & 55        \\
                                            & 50    & 56    & 148       & 105       & 154       & 54        & 66        \\
                                            & 75    & 54    & 176       & 120       & 183       & 61        & 74        \\
                                            & 100   & 59    & 203       & 133       & 210       & 66        & 83        \\ \hline
            \multirow{4}{*}{Approximate}    & 25    & 26    & 60        & 66        & 85        & 22        & 25        \\
                                            & 50    & 29    & 76        & 83        & 114       & 22        & 32        \\
                                            & 75    & 29    & 89        & 86        & 129       & 23        & 32        \\
                                            & 100   & 30    & 104       & 94        & 149       & 23        & 33        \\ \bottomrule
        \end{tabular}
        \vspace{-2.0mm}
    \end{center}
\end{table*}
\begin{table*}[!t]
    \begin{center}
        \caption{Recall of A$k$NN search}   \label{table_aknn-accuracy}
        \vspace{-3.0mm}
        \begin{tabular}{l|lcccccc} \toprule
            \multicolumn{2}{l}{$k$}                 & Crime & HEPMASS   & Household & PAMAP2    & Places    & Wisdom    \\ \midrule
            \multirow{2}{*}{25}         & Average   & 0.86  & 0.91      & 0.84      & 0.80      & 0.91      & 0.85      \\ 
                                        & Median    & 1.00  & 1.00      & 0.96      & 1.00      & 1.00      & 1.00      \\ \hline
            \multirow{2}{*}{50}         & Average   & 0.8   & 0.93      & 0.93      & 0.80      & 0.88      & 0.90      \\
                                        & Median    & 1.00  & 1.00      & 1.00      & 1.00      & 1.00      & 1.00      \\ \hline
            \multirow{2}{*}{75}         & Average   & 0.89  & 0.94      & 0.86      & 0.81      & 0.88      & 0.88      \\
                                        & Median    & 1.00  & 1.00      & 0.93      & 1.00      & 1.00      & 1.00      \\ \hline
            \multirow{2}{*}{100}        & Average   & 0.91  & 0.95      & 0.88      & 0.81      & 0.88      & 0.88      \\
                                        & Median    & 1.00  & 1.00      & 0.94      & 1.00      & 1.00      & 1.00      \\ \bottomrule
        \end{tabular}
        \vspace{-2.0mm}
    \end{center}
\end{table*}

\subsection{Approximate $k$-NN Search}
Last, we apply our distance estimation to approximate $k$-NN search.
In traditional $k$-NN search algorithms, the threshold is initialized at $\infty$, and it is updated during the search.
If we can use a threshold that is (very) close to the exact $k$-NN distance at initialization, the search efficiency is accelerated (although it can produce an approximate result when it is smaller than the exact value).
We tested this idea by using the datasets in Section \ref{section_experiment}.
For each dataset, we randomly sampled 10,000 points and used them as queries.
We initialized the threshold by our estimation and ran $k$-NN search on a $k$d-tree.

Table \ref{table_aknn-time} shows the running time (inference time is also included) with $k = 25$, $50$, $75$, and $100$.
We achieved 1.2x--2.9x speed-up, compared with the exact retrieval.
This result confirms that (i) the above idea is effective and (ii) our estimation does not yield a loose threshold. 

Table \ref{table_aknn-accuracy} shows the average and median recall of the approximate $k$-NN search results.
The average recall is generally high but is less than 1.
This means that, for some queries, the estimated thresholds were less than the exact values, yielding answer sets with less than $k$ points.
However, this is easily alleviated.
For example, if applications require 25 nearest points, they can specify $k$ that is larger than 25 (e.g., $k = 30$), since the time of the estimation-based approximate $k$-NN search does not change much even if $k$ increases a bit.
It is also important to note that the estimation-based approximate $k$-NN search (i) is robust w.r.t. $k$ (which is consistent with the result in Section \ref{section_experiment-result}) and (ii) provides the exact result (with less running time) for many queries, see the median result.

\section{Related Work}  \label{section_related-work}
\textbf{Similarity search} is a primitive operator for many applications, and efficient processing of range and $k$-NN queries has been significantly studied.
Representative approaches are space partitioning ones, such as $k$d-tree and R-tree.
Given a range or $k$-NN query, a branch-and-bound or best-first algorithm is conducted on such a tree to prune unnecessary nodes.
However, when we want to know the distances to the $k$-NNs of the query, these approaches are not efficient, since they involve many data accesses.
For example, $k$d-trees incur $O(kn^{1-1/d})$ data accesses to retrieve $k$-NNs \cite{toth2017handbook}.

Recently, a concept ``index as a model'' \cite{kraska2018case} has been introduced to build indices of datasets, and \textit{learned} indices have been devised.
The first learned index RMI \cite{kraska2018case} was considered for one-dimensional data to alleviate the traversal costs of B+-trees.
An RMI has some neural network models that estimate the position of a given query, and this approach can skip many data accesses.
Some works introduced learned indices to multi-dimensional data \cite{ding2020tsunami, li2020lisa, nathan2020learning, qi2020effectively, zhang2021sprig}.
However, their objectives are different from ours or their adaptability is limited.
For example, literature \cite{zhang2021sprig} considered only 2-dimensional data.
Literatures \cite{ding2020tsunami, nathan2020learning} assume multi-attribute data and focus on relational range queries that search $x$ satisfying $(a \leq x[1] \leq b ) \wedge \cdots \wedge (a' \leq x[d] \leq b')$, where $x[i]$ shows the $i$-th attribute of $x$.
Although \cite{li2020lisa, qi2020effectively} support (approximate) $k$-NN queries, their objective is to minimize disk I/O cost and the indices are specific to disk blocks.
Their approaches are trivially not suitable for in-memory distance estimation, and Pivot in Section \ref{section_experiment} is a better competitor as it needs only $O(1)$ time for our problem.

\vsp
\noindent
\textbf{Cardinality/Selectivity estimation} is also related to our work.
In a nutshell, this problem estimates the output size of a range query.
Query cost estimation and making a batch query processing schedule receive benefits from this problem. 
A classic approach that tries to accurately estimate the cardinality of a given query is multi-dimensional histogram \cite{bruno2001stholes, ioannidis2003history}.
Given a query and a threshold, this approach finds the bin that corresponds to the query.
Assuming that points in this bin follows a specific (e.g., uniform) distribution, this approach estimates the cardinality.
However, real datasets generally do not follow a specific distribution, so the accuracy of this approach tends to be low.

To effectively adapt to real data distributions, ML-based cardinality estimation techniques were devised, e.g., in \cite{DBLP:conf/dolap/JiA0H22, liu2021lhist, sun2021learned, zhu2021flat}.
Some works, e.g., \cite{dutt2019selectivity}, employ decision-tree-based models \cite{chen2016xgboost}.
(Section \ref{section_experiment} shows that PivNet beats the state-of-the-art decision-tree-based model.)
Neural network models were also considered for cardinality estimation \cite{sun2021learned, wang2020monotonic, wang2021consistent, zhu2021flat}, but they cannot provide multiple values as the output at a time.
Moreover, it is not trivial to employ these cardinality estimation models in our problem.

\section{Conclusion}    \label{section_conclusion}
Motivated by the importance of scaling proximity-based data analysis, we addressed the problem of $k$-NN distance estimation.
As a solution for this problem, we proposed a machine-learning technique, PivNet.
This is a neural network-based regression model that estimates the distances to the $k$-NNs of a given query at a time in $O(1)$ time.
This model is trained not only from training queries but also from their nearest pivots to make it more accurate.
We conducted experiments on real datasets.
The results demonstrate that PivNet yields a small estimation error while keeping fast inference time.
Moreover, we did case studies to investigate the usefulness of our model for existing proximity-based data analysis tools.
Thanks to its accuracy and efficiency, we observed that it provides faster analysis with high accuracy.

This paper focused on low-dimensional data.
It is natural to use dimensionality reduction for high-dimensional data cases, such as principal component analysis and t-SNE \cite{maaten2008visualizing}, in order to fit them into our assumption.
However, dimensionality reduction may break the original global distribution, so there would exist applications that need to deal with high-dimensional data as it is.
Estimation of $k$-NN distances for high-dimensional data is an open problem.

\begin{acks}
This research is partially supported by JSPS Grant-in-Aid for Scientific Research (A) Grant Number 18H04095, JST PRESTO Grant Number JPMJPR1931, and JST CREST Grant Number JPMJCR21F2.
\end{acks}

\bibliographystyle{ACM-Reference-Format}
\bibliography{acmart}

\end{document}